\documentclass[onecolumn,10pt,twoside]{svjour3}

\usepackage{subcaption}
\usepackage{hyperref}
\usepackage{graphicx}
\usepackage{listings}
\usepackage{xcolor}
\usepackage{algorithm, algorithmicx, algpseudocode}
\usepackage{float}
\usepackage{enumitem}

\def\makeheadbox{\relax}

\date{}

\begin{document}

\title{SHARI -- An Integration of Tools to Visualize the Story of the Day}

\author{Shawn M. Jones \and Alexander C. Nwala \and Martin Klein \and 
Michele C. Weigle \and Michael L. Nelson}
% \email{{smjones,mklein}@lanl.gov}
\institute{
  Shawn M. Jones \and Martin Klein \at Los Alamos National Laboratory, Los Alamos, NM
  \and
  Alexander C. Nwala \and Michele C. Weigle \and Michael L. Nelson \at Old Dominion University, Norfolk, VA
}

\maketitle

\begin{abstract}
  Tools such as Google News and Flipboard exist to convey daily news, but what about the past? In this paper, we describe how to combine several existing tools with web archive holdings to perform news analysis and visualization of the ``biggest story'' for a given date. StoryGraph clusters news articles together to identify a common news story. Hypercane leverages ArchiveNow to store URLs produced by StoryGraph in web archives. Hypercane analyzes these URLs to identify the most common terms, entities, and highest quality images for social media storytelling. Raintale then takes the output of these tools to produce a visualization of the news story for a given day. We name this process SHARI (StoryGraph Hypercane ArchiveNow Raintale Integration). 
  % This general method can be applied to anything to measure linkage between sources.
\end{abstract}

% \keywords{news, web archives, memento, storytelling, visualization, summarization}

\section{Introduction}

\begin{figure}[htbp]
  \begin{lstlisting}[basicstyle=\tiny]
    {
    "config": "/files/config/polar-media-consensus-graph/f6e84be9969ecef7adb20689002608d0/",
    "connected-comps": [
      {
        "avg-degree": 4.318181818181818,
        "density": 0.10042283298097252,
        "node-details": {
          "annotation": "polarity",
          "color": "green",
          "connected-comp-type": "event"
        },
        "nodes": [
          0,
          1,
  ... additional node ids omitted for brevity ...
        ],
        "unique-source-count": 14
      },
      {
        "avg-degree": 1,
        "density": 1,
        "node-details": {
          "annotation": "polarity",
          "color": "red",
          "connected-comp-type": "cluster"
        },
        "nodes": [
          9,
          67
        ],
        "unique-source-count": 2
      }
    ],
    "links": [
      {
        "rank": 1,
        "sim": 0.57,
        "source": 2,
        "target": 21,
        "label": "1 (0.57)",
        "label-description": "rank (sim)"
      },
  ... additional link definitions omitted for brevity ...
      {
        "rank": 96,
        "sim": 0.3,
        "source": 53,
        "target": 73,
        "label": "96 (0.3)",
        "label-description": "rank (sim)"
      }
    ],
    "ner-version": "3.8.0",
    "nodes": [
  ... other nodes omitted for brevity ...
      {
        "entities": [
          {
            "class": "LOCATION",
            "entity": "Coney Island"
          },
          {
            "class": "LOCATION",
            "entity": "Brooklyn"
          },
          {
            "class": "PERSON",
            "entity": "Victor J. Blue"
          },
  ...
  
        ],
        "extraction-time": "2020-03-23T00:09:10.325362",
        "favicon": "https://www.nytimes.com/vi-assets/static-assets/favicon-4bf96cb6a1093748bf5b3c429accb9b4.ico",
        "id": "nytimes.com-1",
        "link": "https://www.nytimes.com/2020/03/22/health/coronavirus-restrictions-us.html",
        "node-details": {
          "annotation": "polarity",
          "color": "blue",
          "connected-comp-type": "event",
          "type": "left"
        },
        "published": "Sun, 22 Mar 2020 22:00:52 +0000",
        "rss-uri-m": "https://web.archive.org/web/20200323000609id_/https://rss.nytimes.com/services/xml/rss/nyt/HomePage.xml",
        "text": "Health |Harsh Steps Are Needed to Stop the Coronavirus, Experts Say\nhttps://nyti.ms/3dkfoCc\nA beach stroller in the Coney Island neighborhood of Brooklyn on Saturday.Credit...Victor J. Blue for The New York Times\nHarsh Steps Are Needed to Stop the Coronavirus, Experts Say\nScientists who have fought pandemics describe difficult measures needed to defend the United States against a fast-moving pathogen.\nA beach stroller in the Coney Island neighborhood of Brooklyn on Saturday.Credit...Victor J. Blue for The New York Times\nSupported by\nBy Donald G. McNeil Jr.\nMarch 22, 2020, 6:00 p.m. ET\nTerrifying though the coronavirus may be, it can be turned back. China, South Korea, Singapore and Taiwan have demonstrated that, with furious efforts, the contagion can be brought to heel.\nWhether they can keep it suppressed remains to be seen...",
        "title": "Harsh Steps Are Needed to Stop the Coronavirus, Experts Say - The New York Times"
      },
  ... other articles omitted for brevity ...
    ],
    "self": "http://storygraph.cs.odu.edu/graphs/polar-media-consensus-graph/#cursor=0&hist=1440&t=2020-03-23T00:09:10",
    "timestamp": "2020-03-23T00:09:10.707796Z",
    "graph-pointer": {
      "cursor": 0,
      "hist": 1440,
      "cur-path": "2020/03/23"
    }
  }
  \end{lstlisting}
  \caption{An abridged version of the JSON file generated by StoryGraph that drives the visualization in Figure \ref{fig:storygraph-march-23-2020}.}
  \label{fig:storygraphjson}
  \end{figure}

\begin{figure}[htbp]
  \centering
  \includegraphics[width=\textwidth]{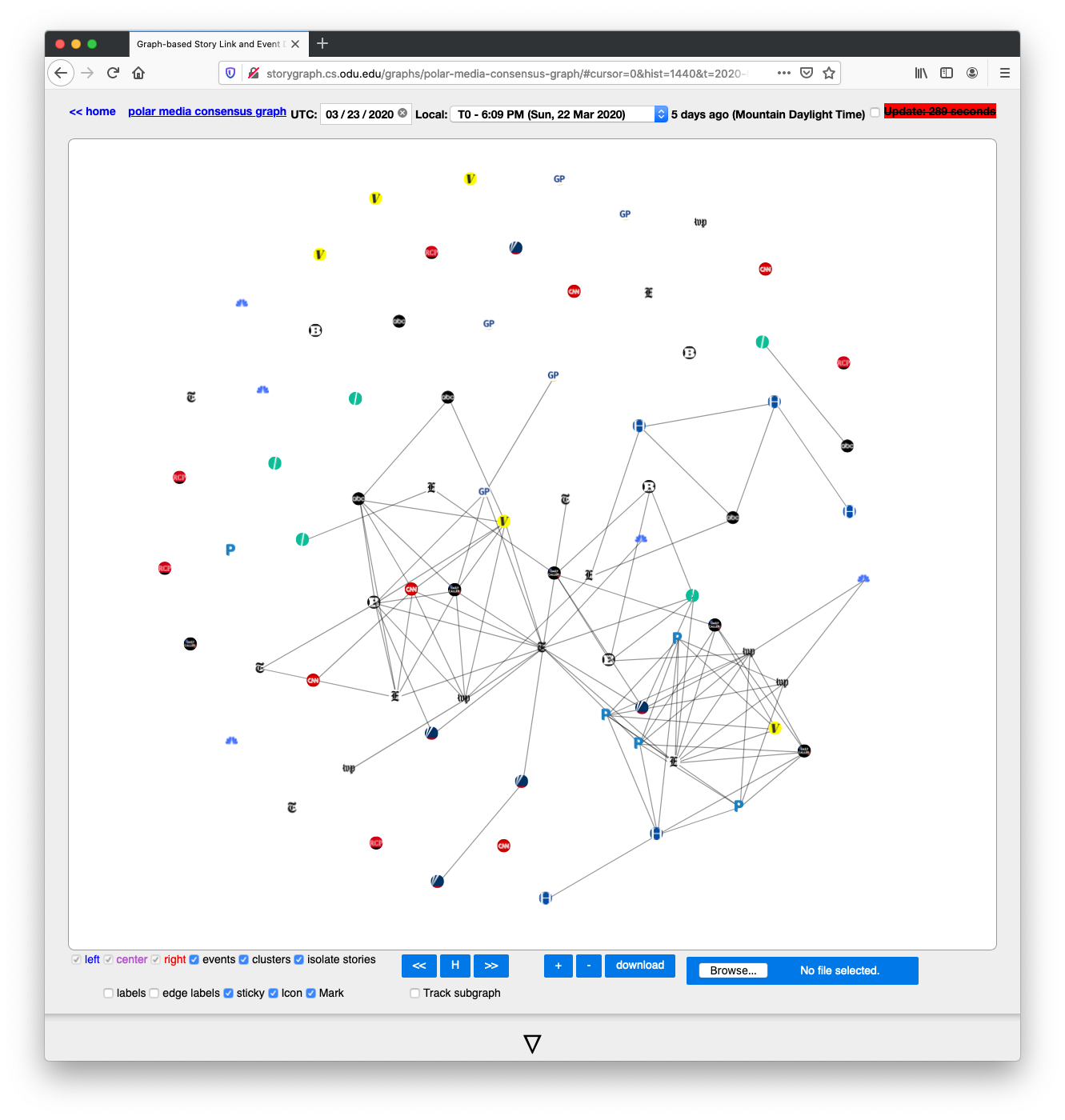}
  \caption{The StoryGraph news similarity graph for March 23, 2020.\\URL:\url{http://storygraph.cs.odu.edu/graphs/polar-media-consensus-graph/\#cursor=0\&hist=1440\&t=2020-03-23T00:09:10}}
  \label{fig:storygraph-march-23-2020}
\end{figure}

\begin{figure}[htbp]
  \centering
  \includegraphics[width=\textwidth]{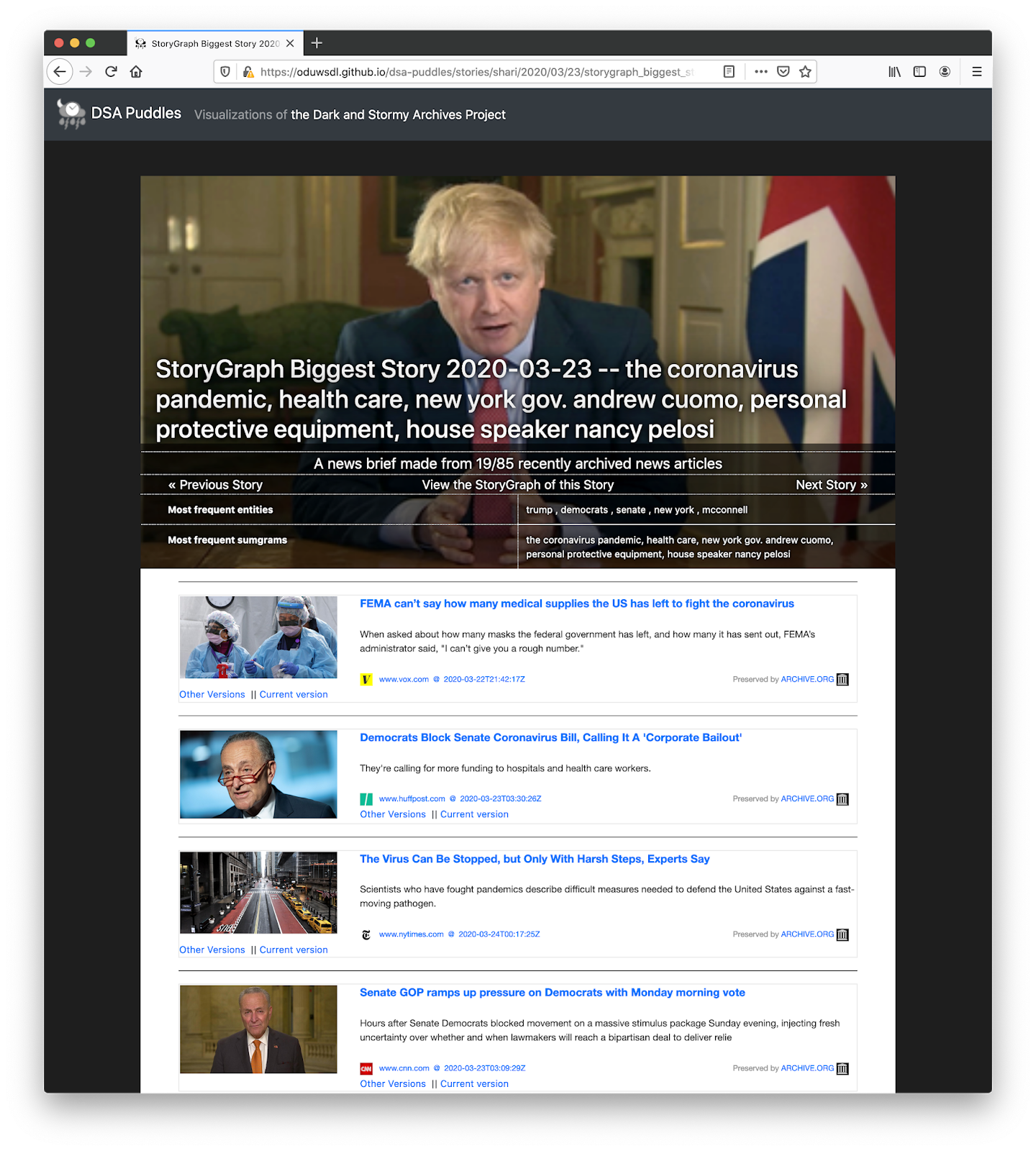}
  \caption[The ``biggest news story'' of March 23, 2020 produced by the SHARI process.]{The ``biggest news story'' of March 23, 2020 produced by the SHARI process.\\URL:\url{https://oduwsdl.github.io/dsa-puddles/stories/shari/2020/03/23/storygraph_biggest_story_2020-03-23/}}
  \label{fig:shari-story-example}
\end{figure}

% \protect\footnotemark
% \footnotetext{}

\begin{figure}[t]
  \centering
  \includegraphics[width=\textwidth]{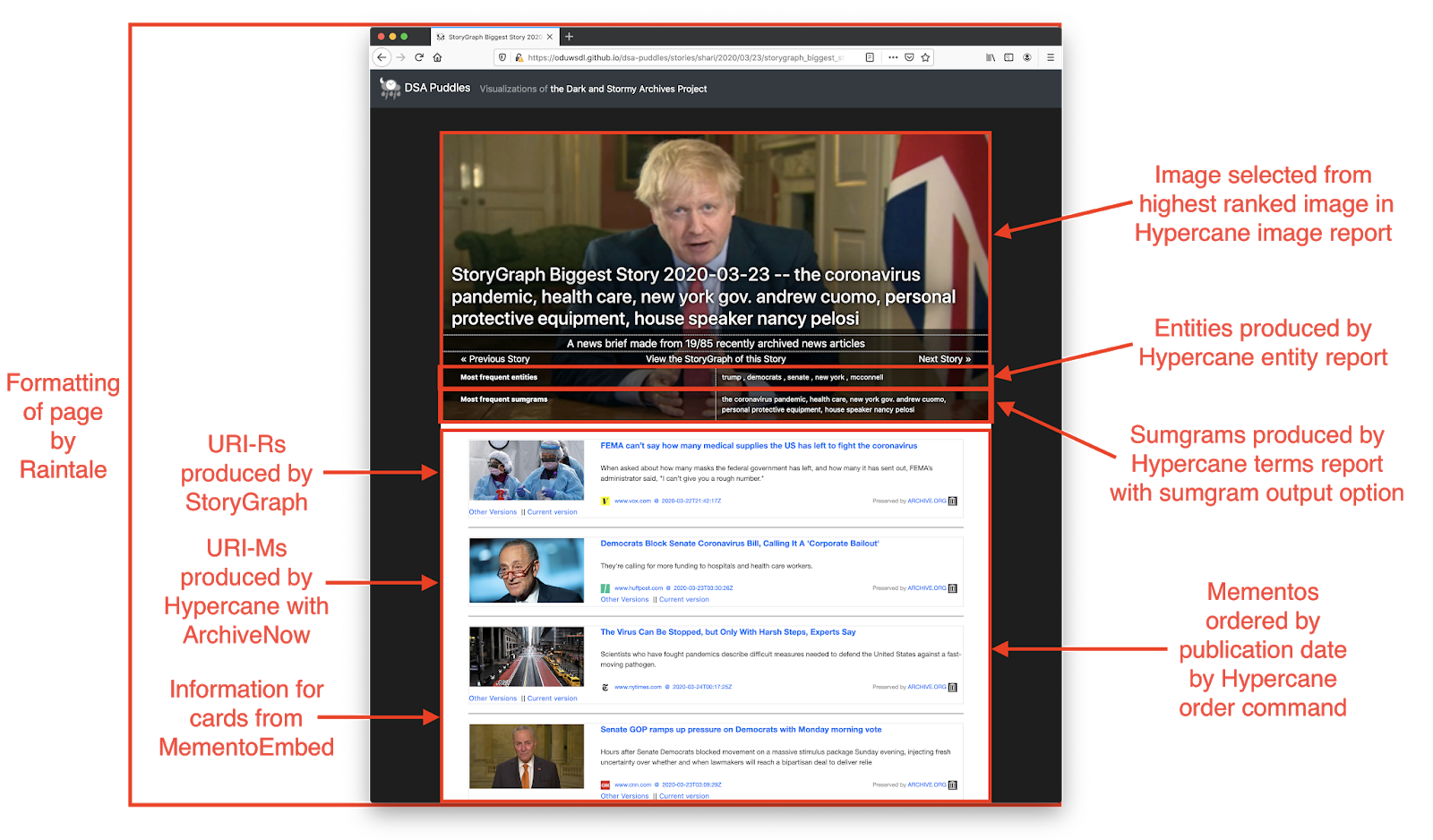}
  \caption{Annotations detail which SHARI components provide each part of the visualization shown in Figure \ref{fig:shari-story-example}.}
  \label{fig:shari-result-annotated}
\end{figure}

\begin{figure}[htbp]
  \centering
  \includegraphics[width=0.95\textwidth]{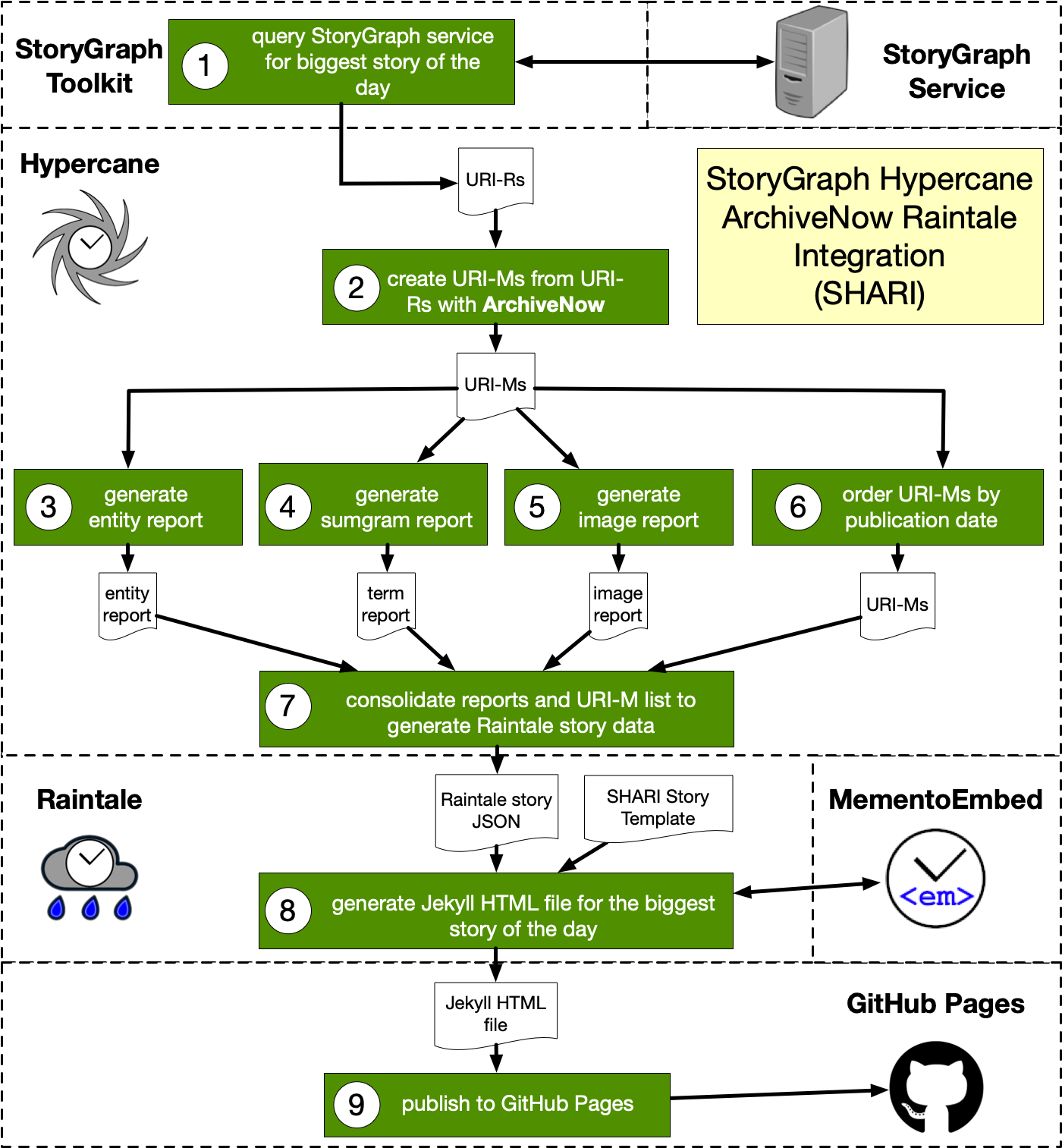}
  \caption{SHARI process for creating a visualization of the biggest news story for a given day}
  \label{fig:shari-process}
\end{figure}

Tools such as Google News and Flipboard exist to convey daily news, but what about the news of the past? We have combined StoryGraph\footnote{\url{http://storygraph.cs.odu.edu}} with tools from the Dark and Stormy Archives Toolkit\footnote{\url{https://oduwsdl.github.io/dsa/software.html}} to produce the StoryGraph Hypercane ArchiveNow Raintale Integration (SHARI) process. These tools represent disparate research efforts in news analysis, corpus summarization, web archiving, and visualization. The integration produces a summary of the ``biggest story'' for a given date. SHARI combines the following components from Old Dominion University's Web Science and Digital Libraries Research Group\footnote{\url{https://ws-dl.cs.odu.edu}}:
\begin{itemize}
  \item StoryGraph: a platform that downloads RSS feeds and analyzes the linked articles to cluster news stories \cite{nwala_365_dots_2020} -- \url{http://storygraph.cs.odu.edu/}
  \item Hypercane: a framework for intelligently sampling and analyzing documents from web archive collections \cite{jones_hypercane_2020} -- \url{https://oduwsdl.github.io/hypercane}
  \item ArchiveNow: a library developed by Aturban et al. \cite{aturban_archivenow:_2018} that submits live web URI-Rs to web archives to create URI-Ms -- \url{https://github.com/oduwsdl/archivenow}
  \item Raintale: a MementoEmbed \cite{jones_preview_2018} client that creates stories from a sample of mementos -- \url{https://oduwsdl.github.io/raintale}
\end{itemize}

Nwala et al. \cite{nwala_bootstrapping_2018,nwala_scraping_2018} have focused on finding seeds within search engine result pages (SERPs), social media stories, and news feeds. As part of this research, Nwala et al. also developed StoryGraph \cite{nwala_365_dots_2020}, a service that saves RSS feeds from 17 news sources (Table \ref{tab:17_news_sources} in Appendix A) every ten minutes. With these RSS feeds, StoryGraph analyzes the lexical connections between articles across feeds to generate JSON output, which drives a graph visualization. Figure \ref{fig:storygraphjson} displays some of this JSON output for March 23, 2020. StoryGraph then visualizes this output, as shown in Figure \ref{fig:storygraph-march-23-2020}.

Collections on specific topics exist at various web archives \cite{jones_many_2018}. AlNoamany et al. \cite{alnoamany_generating_2017} introduced how to use social media storytelling to summarize web archive collections. Klein et al. \cite{10.1145/3201064.3201085} have built collections from web archives by conducting focused crawls. Jones developed Hypercane \cite{jones_hypercane_2020} to intelligently sample mementos from larger collections. Jones also developed Raintale \cite{jones_raintale_2019} for generating social media stories to summarize groups of mementos, providing visualizations that employ familiar techniques, like cards, that require no training for most users to understand.

The JSON data structure from Figure \ref{fig:storygraphjson} provides all information gathered but is difficult for humans to understand at a glance. The graph shown in Figure \ref{fig:storygraph-march-23-2020} provides an overview of the JSON through favicons and edges, but a user requires some training to fully comprehend what it represents. Figure \ref{fig:shari-story-example} displays the largest connected component from this graph visualized via the SHARI process. Through images, text snippets, titles, cards, domain names, favicons, and other content, the SHARI output allows the viewer to intuitively understand that the biggest news story for this date consists of different reactions to the growing COVID-19 pandemic.

\section{The SHARI process}

The StoryGraph Hypercane ArchiveNow Raintale Integration (SHARI) \cite{jones_shari_2020} process automatically creates stories summarizing news for a day. Figure \ref{fig:shari-result-annotated} details what each tool contributes to the story. Figure \ref{fig:shari-process} shows the steps of the SHARI process. 

\begin{figure}[htbp]
  \centering
  \includegraphics[width=0.5\textheight]{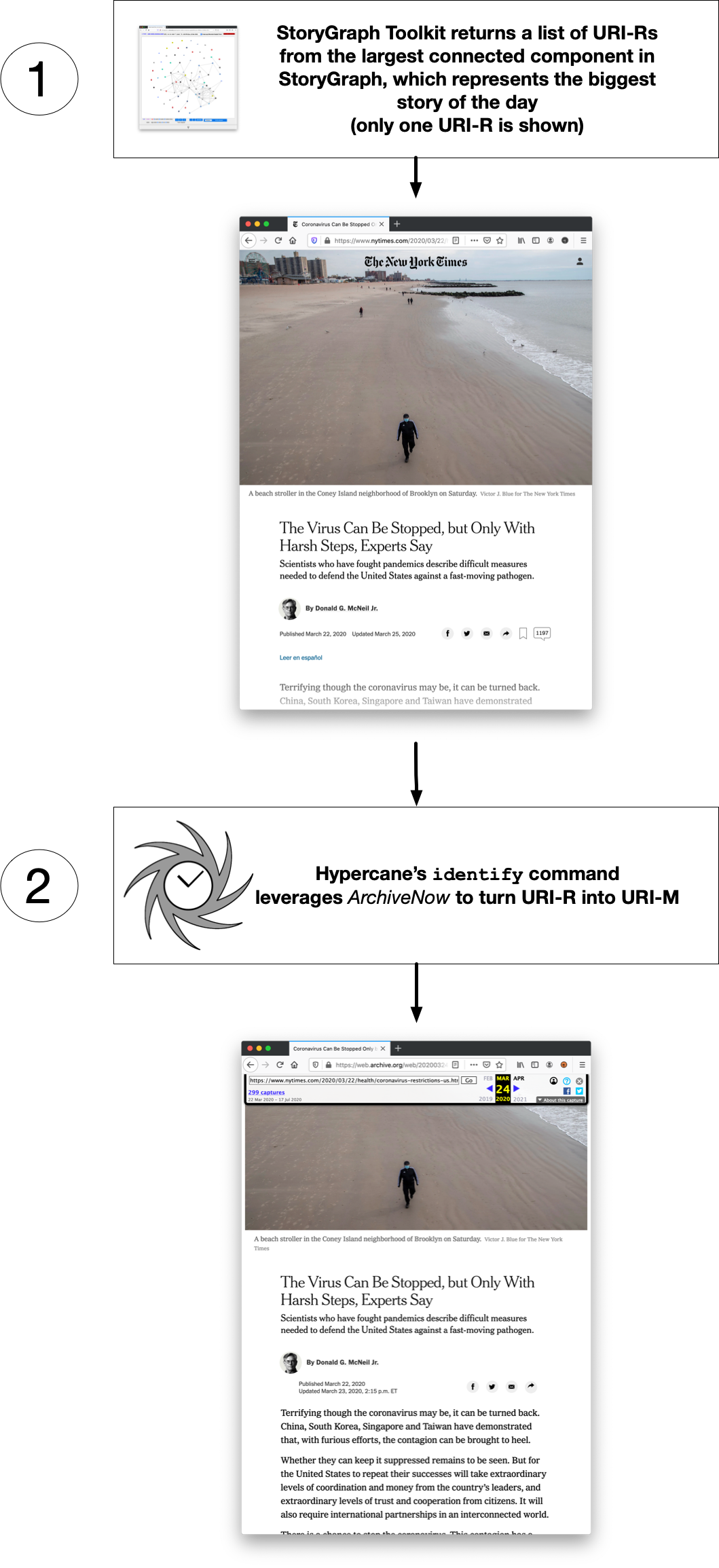}
  \caption{SHARI steps 1-2 illustrated with a \emph{single URI-R} from the story shown in Figure \ref{fig:shari-story-example}. Here SHARI extracts the URI-R from StoryGraph and then creates a corresponding URI-M with ArchiveNow.}
  \label{fig:shari-steps12-single-urir}
\end{figure}

\begin{figure}[htbp]
  \centering
  \includegraphics[width=0.95\textwidth]{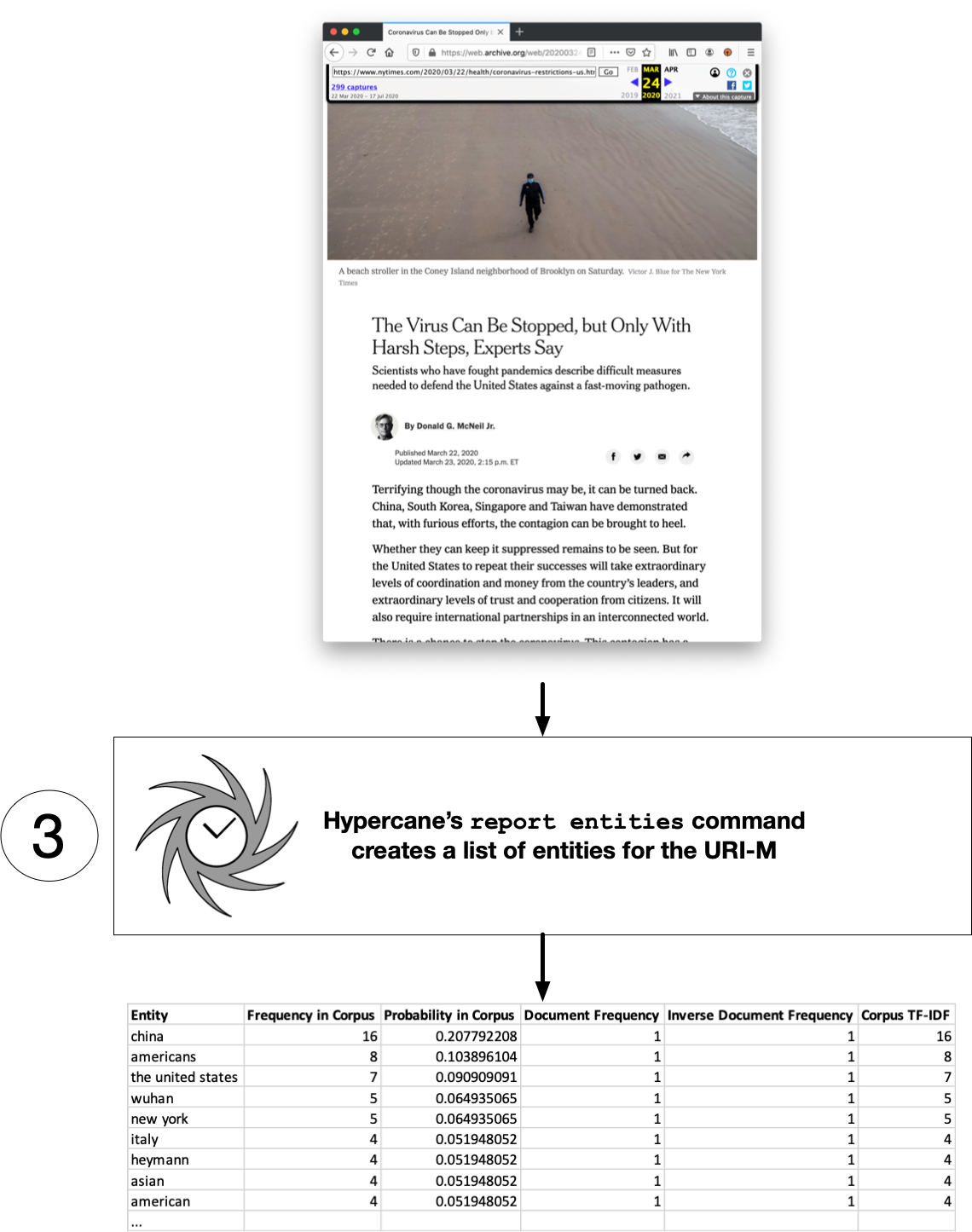}
  \caption{SHARI step 3 reproting entities from the URI-M generated in Figure \ref{fig:shari-steps12-single-urir}}
  \label{fig:shari-step3-single-urir}
\end{figure}

\begin{figure}[htbp]
  \centering
  \includegraphics[width=0.95\textwidth]{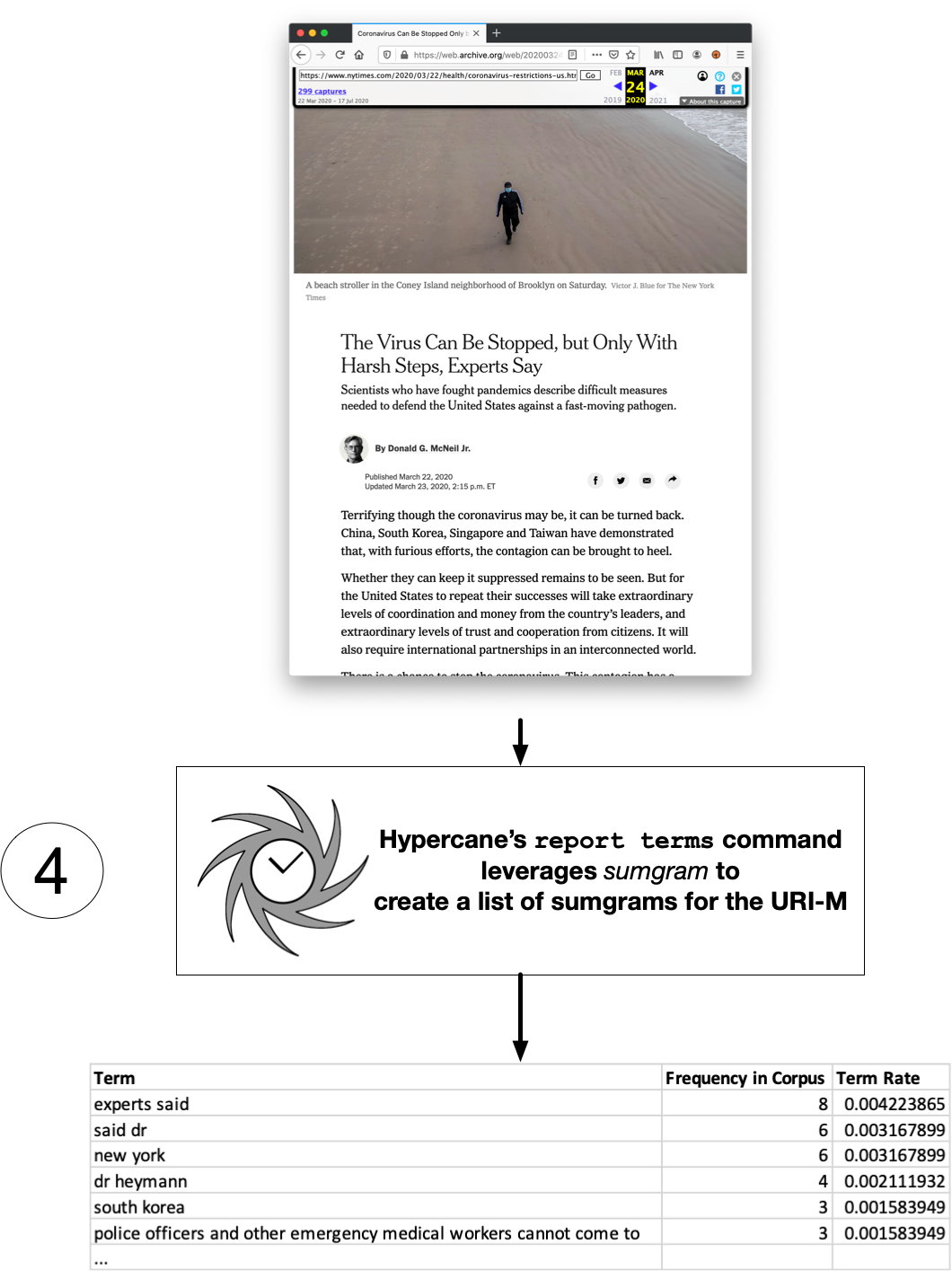}
  \caption{SHARI step 4 reporting sumgrams from the URI-M generated in Figure \ref{fig:shari-steps12-single-urir}}
  \label{fig:shari-step4-single-urir}
\end{figure}

\begin{figure}[htbp]
  \centering
  \includegraphics[width=0.95\textwidth]{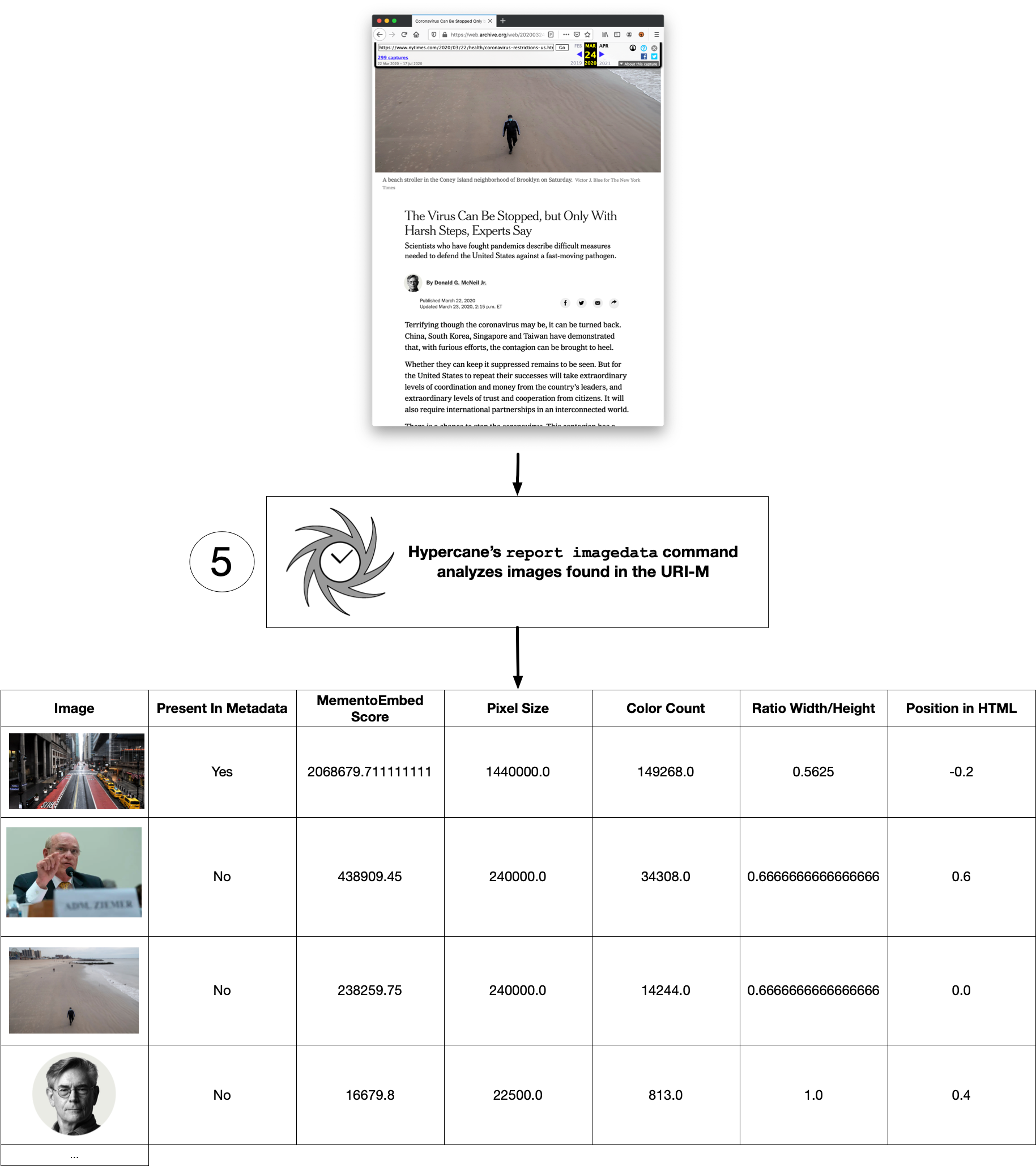}
  \caption{SHARI step 5 reporting a image metrics from the URI-M generated in Figure \ref{fig:shari-steps12-single-urir}}
  \label{fig:shari-step5-single-urir}
\end{figure}

\begin{figure}[htbp]
  \centering
  \includegraphics[width=0.95\textwidth]{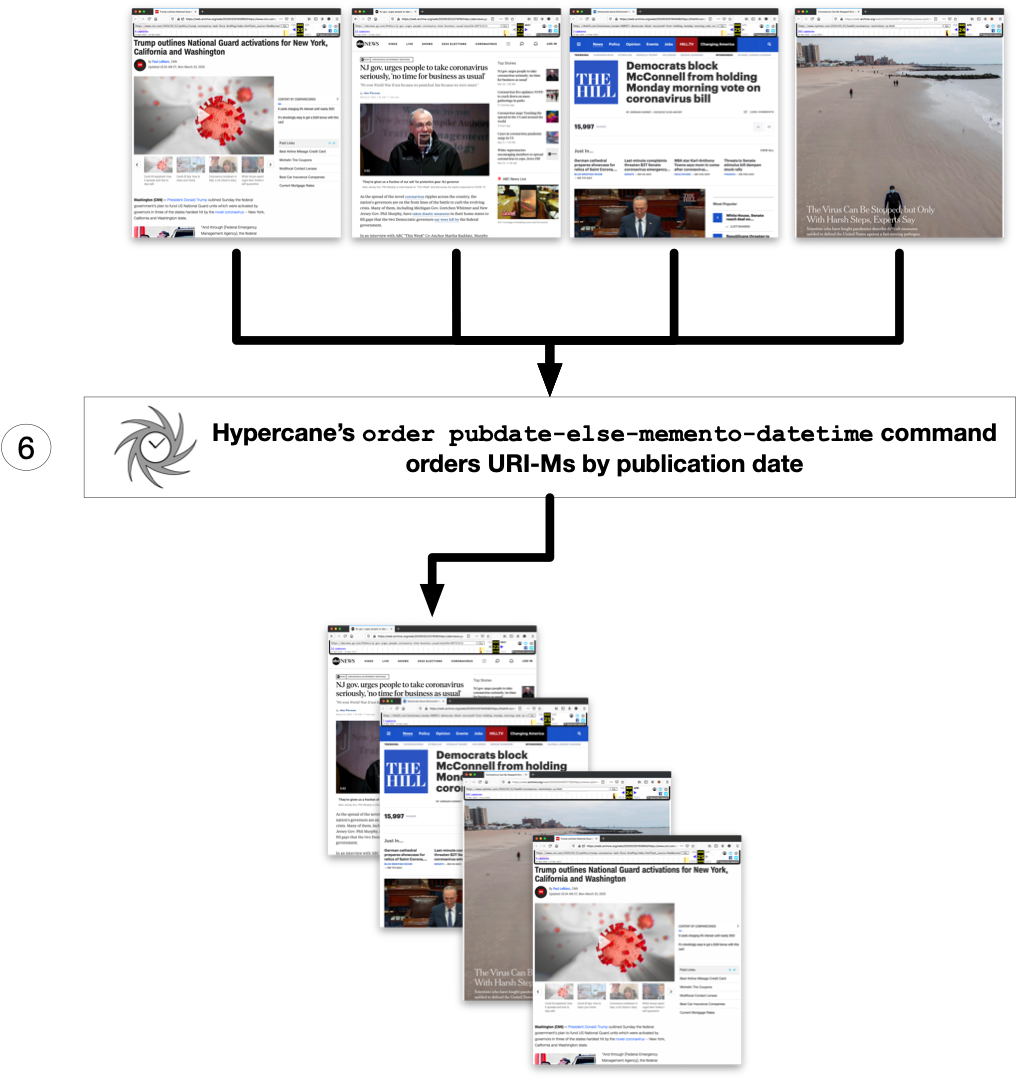}
  \caption{SHARI step 6 orders all mementos first by publication date, then memento-datetime.}
  \label{fig:shari-step6}
\end{figure}

\begin{figure}[htbp]
  \centering
  \includegraphics[width=0.95\textwidth]{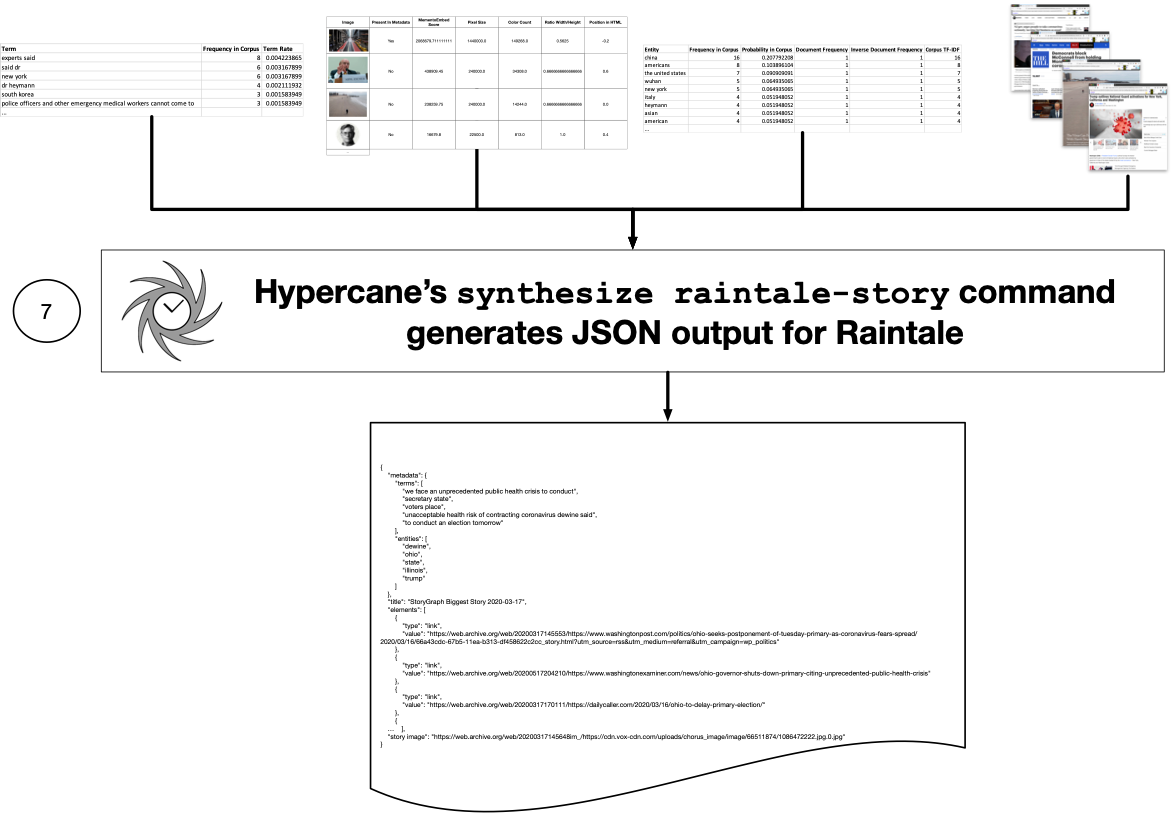}
  \caption{SHARI step 7 combines all data into a JSON format used by Raintale for storytelling.}
  \label{fig:shari-step7}
\end{figure}

\begin{figure}[htbp]
  \centering
  \includegraphics[width=0.75\textwidth]{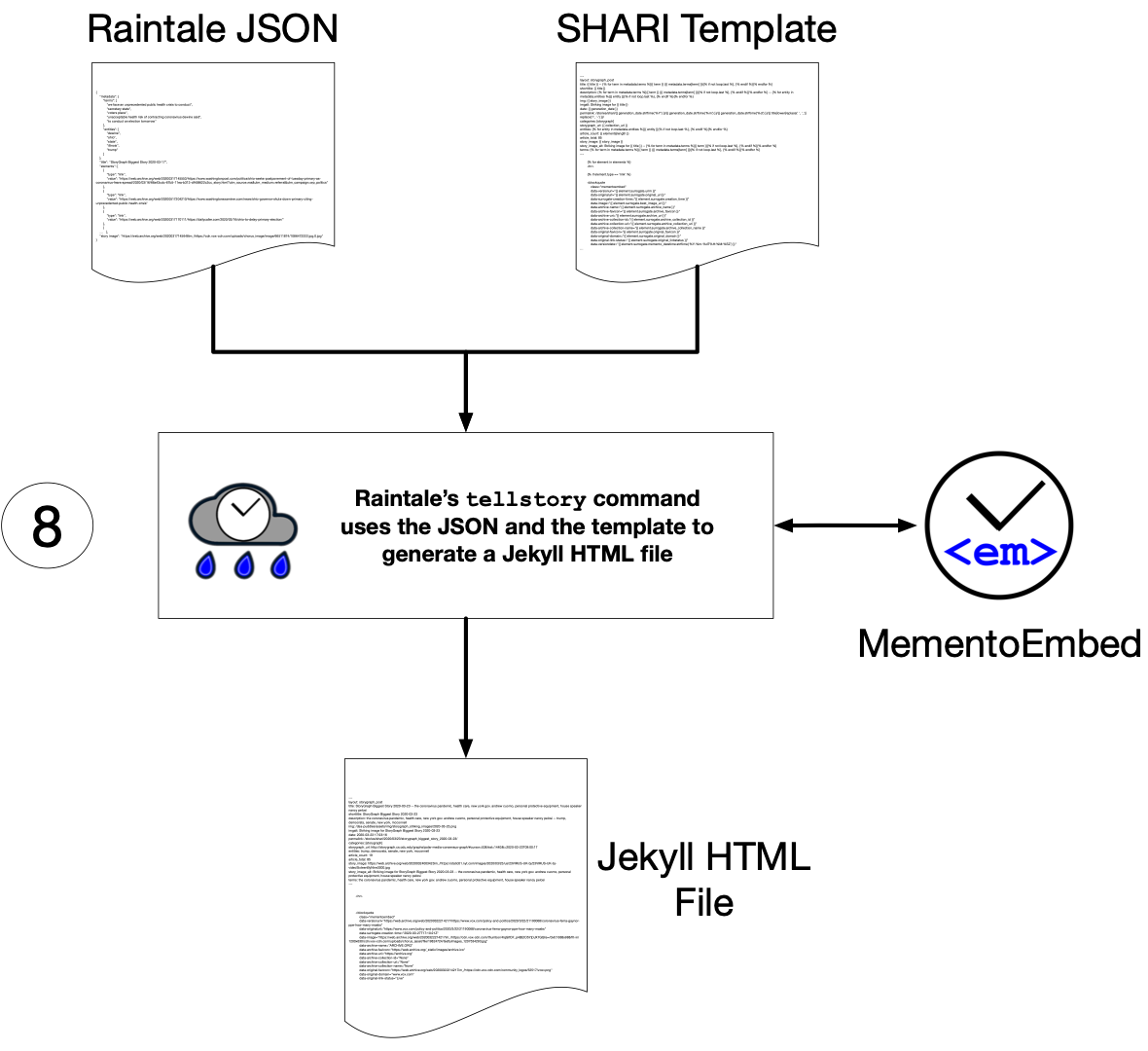}
  \caption{SHARI step 8 feeds the JSON file from Step 7 and a template file into Raintale to generate the story. Raintale queries MementoEmbed for information about each memento.}
  \label{fig:shari-step8}
\end{figure}

\begin{figure}
  \centering
  \includegraphics[width=0.6\textwidth]{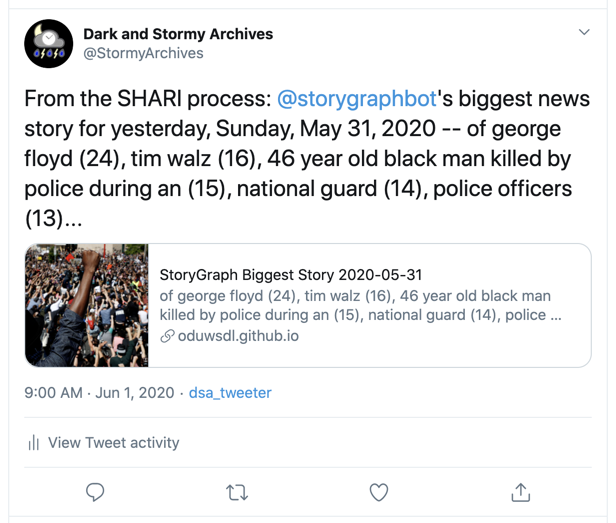}
  \caption{The \emph{dsa\_tweeter} bot announces the availability of new SHARI stories each day.}
  \label{fig:daily-tweet}
\end{figure}

\begin{enumerate}
\item With the StoryGraph Toolkit, we query the StoryGraph service for the URI-Rs belonging to the biggest story of the day.
\item Hypercane converts these URI-Rs to URI-Ms by first attempting to find a corresponding URI-M by querying the LANL Memento Aggregator\footnote{\url{https://timetravel.mementoweb.org}} via the Memento Protocol \cite{van_de_sompel_rfc_2013}. For each URI-M that does not have a memento, Hypercane creates a memento by calling ArchiveNow \cite{aturban_archivenow:_2018} (Figure \ref{fig:shari-steps12-single-urir}).
\item Hypercane runs the mementos through spaCy\footnote{\url{https://spacy.io/}} to generate a list of named entities, sorted by frequency (Figure \ref{fig:shari-step3-single-urir}).
\item Hypercane runs the mementos through sumgram \cite{nwala_sumgram_2019} and generates a list of sumgrams, sorted by frequency (Figure \ref{fig:shari-step4-single-urir}).
\item Hypercane scores all of the mementos' embedded images. Images that article authors reference in HTML META tags are favored first, followed by MementoEmbed \cite{jones_preview_2018} score, then pixel size, color count, the ratio of width to height, and finally position on the page (Figure \ref{fig:shari-step5-single-urir}).
\item Hypercane runs the mementos through newspaper3k\footnote{\url{https://newspaper.readthedocs.io/en/latest/}} to extract each article's publication date and orders the URI-Ms by that date (Figure \ref{fig:shari-step6}) .
\item Hypercane consolidates the entities, terms, image scores, and ordered URI-Ms into a JSON file containing the structured data for the summary. During this step, Hypercane uses the highest scoring image as the striking image for the summary (Figure \ref{fig:shari-step7}). In Figure \ref{fig:shari-result-annotated}, the highest-ranking image is the UK Prime Minister addressing his country about the COVID-19 pandemic. 
\item Raintale renders the output as Jekyll HTML based on the contents of this JSON file, a template file, and information on each memento provided by MementoEmbed (Figure \ref{fig:shari-step7}).
\item The SHARI script publishes the summary story to GitHub Pages for distribution. Figure \ref{fig:daily-tweet} shows the output of our \emph{dsa\_tweeter} bot which announces the story after publication through the \emph{@StormyArchives} Twitter account.
\end{enumerate}

\section{Discussion}

\begin{figure}[htbp]
  \centering
  \includegraphics[width=0.95\textwidth]{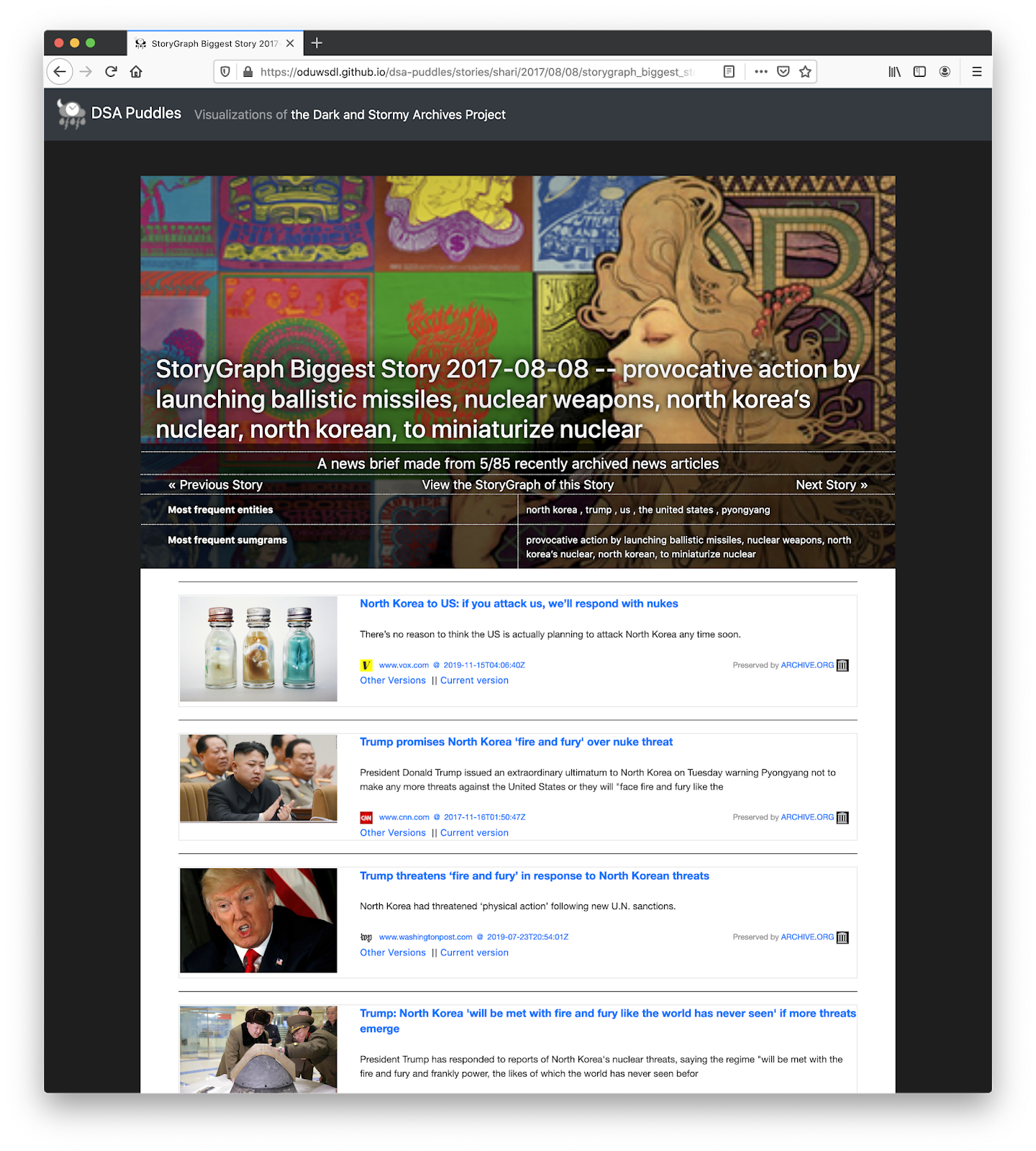}
  \caption{SHARI output for August 8, 2017 - the launch date of StoryGraph\\URL: \url{https://oduwsdl.github.io/dsa-puddles/stories/shari/2017/08/08/storygraph_biggest_story_2017-08-08/}}
  \label{fig:shari-process-2017-08-08}
\end{figure}

\begin{figure}[htbp]
  \centering
  \includegraphics[width=0.95\textwidth]{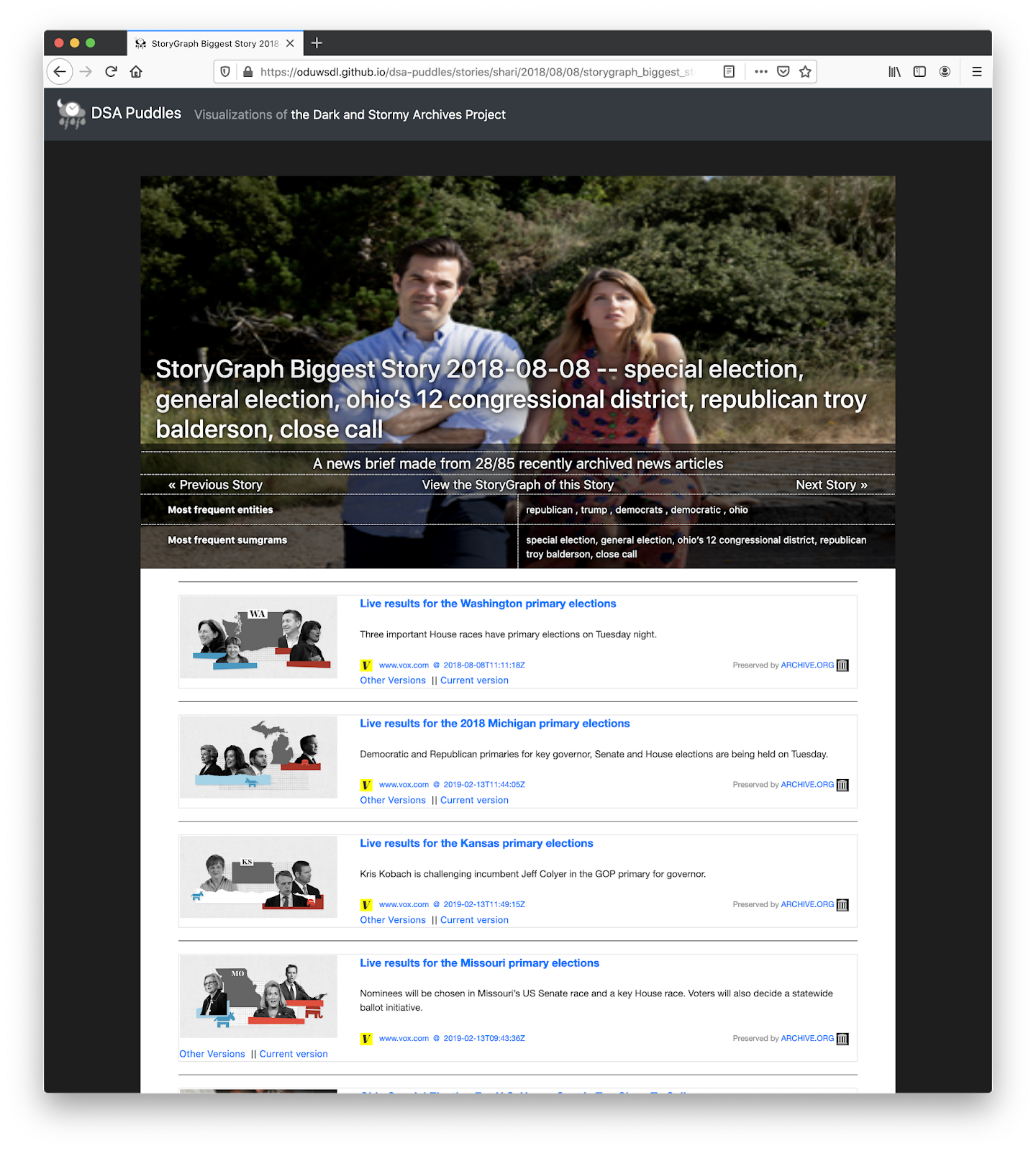}
  \caption{SHARI output for August 8, 2018 - a year after the launch date of StoryGraph\\URL: \url{https://oduwsdl.github.io/dsa-puddles/stories/shari/2018/08/08/storygraph_biggest_story_2018-08-08/}}
  \label{fig:shari-process-2018-08-08}
\end{figure}

\begin{figure}[htbp]
  \centering
  \includegraphics[width=0.95\textwidth]{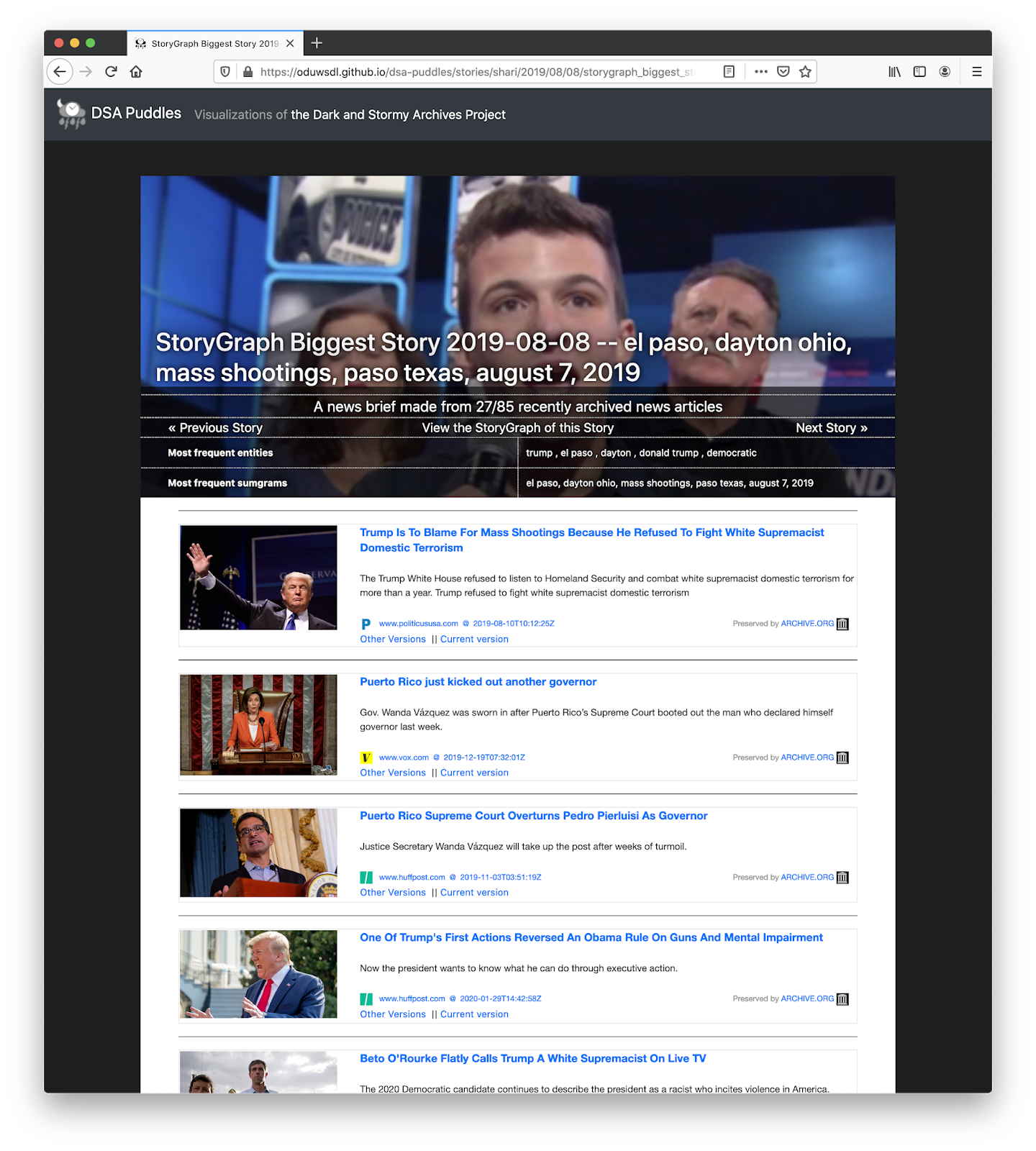}
  \caption{SHARI output for August 8, 2019 - two years after the launch date of StoryGraph\\URL: \url{https://oduwsdl.github.io/dsa-puddles/stories/shari/2019/08/08/storygraph_biggest_story_2019-08-08/}}
  \label{fig:shari-process-2019-08-08}
\end{figure}

StoryGraph is a valuable resource with additional unrealized potential. We are not only able to create stories for today or yesterday but any date back to August 8, 2017, when Nwala launched StoryGraph. As seen in Figures \ref{fig:shari-process-2017-08-08}, \ref{fig:shari-process-2018-08-08}, and \ref{fig:shari-process-2019-08-08} we can see how the world has evolved each year on StoryGraph's launch date. In Figure \ref{fig:shari-process-2017-08-08}, the biggest news story was that of North Korea threatening other nations with nuclear weapons. One year later, in Figure \ref{fig:shari-process-2018-08-08}, we see that the biggest news story is the results of several United States Congressional and gubernatorial primaries. Two years after StoryGraph's launch, Figure \ref{fig:shari-process-2019-08-08} shows that the biggest news story is the aftermath of the 2019 shootings in El Paso and Dayton.

\section{Summary and Future Work}

SHARI produces a familiar yet novel method of viewing news for a given day. SHARI can create stories for today, yesterday, and back to StoryGraph's creation on August 8, 2017. It is different from other storytelling services like Wakelet\footnote{\url{https://wakelet.com/}} because SHARI is entirely automated. The stories produced by SHARI are different from services like Google News\footnote{\url{https://news.google.com/}} or Flipboard\footnote{\url{https://flipboard.com/}} because those tools focus on current events and personalized topics. Because StoryGraph samples content from multiple sides of the political spectrum, the SHARI process can provide a visualization of articles not tied to one interest area or even a single side's terminology. This process works because each component is loosely coupled, has high cohesion, has explicit interfaces, and engages in information hiding. Each command passes data in the expected format to the next.  

We are also exploring how to improve striking image selection for stories. One could use this to consider how the same story is told in different venues. For instance, one could ask StoryGraph only to include left-leaning sources and produce a SHARI story. One could then do the same for only the right-leaning sources. With both stories, one could compare the striking images and sumgrams that SHARI produces. We are investigating how to produce and render other news stories for a given day and any given period of time. Finally, we are examining how to best visualize significant events that span substantial periods of time, like the entire COVID-19 news story.  Though StoryGraph is an existing service that gathers current news, we also want to apply its algorithm directly to mementos and tell the news stories of past events like the Hurricane Katrina disaster.

\section{Acknowledgements}

This work supported in part by the Institute of Museum and Library Services (LG-71-15-0077-15).

\bibliographystyle{acm}
\bibliography{references}

\clearpage

\section{Appendix A: StoryGraph News Sources}

\begin{table}[htbp]
  \caption{The 17 news sources analyzed by StoryGraph}
  \label{tab:17_news_sources}
  \begin{tabular}{l|l|l}
  \textbf{News Source} & \textbf{Feed URL}                                        & \textbf{US Political Polarity} \\ \hline
  Politicus USA        & http://www.politicususa.com/feed                         & Left                           \\
  Vox                  & https://www.vox.com/rss/index.xml                        & Left                           \\
  Huffington Post      & http://www.huffingtonpost.com/section/front-page/feed    & Left                           \\
  MSNBC                & http://www.msnbc.com/feeds/latest                        & Left                           \\
  New York Times       & http://rss.nytimes.com/services/xml/rss/nyt/HomePage.xml & Left                           \\
  Washington Post      & http://feeds.washingtonpost.com/rss/politics             & Left                           \\
  CNN                  & http://rss.cnn.com/rss/cnn\_topstories.rss               & Center                         \\
  Politico             & http://www.politico.com/rss/politics.xml                 & Center                         \\
  ABC News             & http://abcnews.go.com/abcnews/topstories                 & Center                         \\
  The Hill             & http://thehill.com/rss/syndicator/19109                  & Center                         \\
  Real Clear Politics  & http://feeds.feedburner.com/realclearpolitics/qlMj       & Center                         \\
  Washington Examiner  & http://www.washingtonexaminer.com/rss/news               & Right                          \\
  Fox News             & http://feeds.foxnews.com/foxnews/latest                  & Right                          \\
  Daily Caller         & http://feeds.feedburner.com/dailycaller                  & Right                          \\
  Conservative Tribune & http://conservativetribune.com/feed/                     & Right                          \\
  Breitbart            & http://feeds.feedburner.com/breitbart                    & Right                          \\
  The Gateway Pundit   & http://www.thegatewaypundit.com/feed/                    & Right                         
  \end{tabular}
\end{table}

\end{document}